\documentclass[traditabstract,bibyear]{aa} 

\newcommand{\gtsim}{\protect\raisebox{-0.5ex}{$\:\stackrel{\textstyle >}
        {\sim}\:$}}
\newcommand{\ltsim}{\protect\raisebox{-0.5ex}{$\:\stackrel{\textstyle <}
        {\sim}\:$}}

\usepackage{graphicx}
\usepackage{natbib,twoopt}
\bibpunct{(}{)}{;}{a}{}{,} 
\makeatother

\begin{document}

\title{The GAPS Programme with HARPS-N at TNG}
\subtitle{X. Differential abundances in the XO-2 planet hosting binary \thanks{Based on observations made with the Italian {\it Telescopio Nazionale Galileo} (TNG), operated on the 
island of La Palma by the INAF - {\it Fundaci\'on Galileo Galilei} at the {\it Roche de los Muchachos} Observatory
of the {\it Instituto de Astrof\'isica de Canarias} (IAC) in the framework of the large programme Global 
Architecture of Planetary Systems (GAPS; P.I. A. Sozzetti).}}
   
\author{
K. Biazzo \inst{1} \and R. Gratton \inst{2} \and S. Desidera \inst{2} \and S. Lucatello \inst{2} \and A. Sozzetti \inst{3} \and A. S. Bonomo \inst{3} 
\and M. Damasso \inst{3} \and D. Gandolfi \inst{4,5} \and L. Affer \inst{6} \and C. Boccato \inst{2} \and F. Borsa \inst{7} \and R. Claudi \inst{2} 
\and R. Cosentino \inst{1,8} \and E. Covino \inst{9} \and C. Knapic \inst{10} \and A. F. Lanza \inst{1} \and J. Maldonado \inst{6} \and F. Marzari \inst{11} 
\and G. Micela \inst{6} \and P. Molaro \inst{10} \and I. Pagano \inst{1} \and M. Pedani \inst{8} \and I. Pillitteri \inst{6} \and G. Piotto \inst{2,11} \and 
E. Poretti \inst{7} \and M. Rainer \inst{7} \and N. C. Santos \inst{12,13,14} \and G. Scandariato \inst{1} \and R. Zanmar Sanchez\inst{1}
}

\offprints{K. Biazzo}
\mail{katia.biazzo@oact.inaf.it}

\institute{INAF - Osservatorio Astrofisico di Catania, Via S. Sofia 78, I-95123 Catania, Italy
\and INAF - Osservatorio Astronomico di Padova, Vicolo dell'Osservatorio 5, I-35122 Padova, Italy
\and INAF - Osservatorio Astrofisico di Torino, Via Osservatorio 20, I-10025 Pino Torinese, Italy
\and Dipartimento di Fisica, Universit\`a di Torino, Via P. Giuria 1, I-10125 Torino, Italy
\and Landessternwarte K\"onigstuhl, Zentrum f\"ur Astronomie der Universitat Heidelberg, K\"onigstuhl 12, D-69117 Heidelberg, Germany
\and INAF - Osservatorio Astronomico di Palermo, Piazza del Parlamento 1, I-90134 Palermo, Italy
\and INAF - Osservatorio Astronomico di Brera, Via E. Bianchi 46, I-23807 Merate (LC), Italy
\and Fundaci\'on Galileo Galilei - INAF, Rambla Jos\'e Ana Fernandez P\'erez 7, E-38712 Bre\~na Baja, TF - Spain
\and INAF - Osservatorio Astronomico di Capodimonte, Salita Moiariello 16, I-80131 Napoli, Italy
\and INAF - Osservatorio Astronomico di Trieste, Via Tiepolo 11, I-34143 Trieste, Italy
\and Dipartimento di Fisica e Astronomia Galileo Galilei - Universit\`a di Padova, Vicolo dell'Osservatorio 2, I-35122, Padova, Italy
\and Instituto de Astrof\'{i}sica e Ci\^encias do Espa\c{c}o, Universitade do Porto, CAUP, Rua das Estrelas, 4150-762 Porto, Portugal
\and Centro de Astrof\'{i}sica, Universitade do Porto, Rua das Estrelas, 4150-762 Porto, Portugal
\and Departamento de F\'{i}sica e Astronomia, Facultade de Ci\^encias, Univ. do Porto, Rua do Campo Alegre, s/n, 4169-007 Porto, Portugal
}

\date{Received .../ accepted ...}

\abstract{Binary stars hosting exoplanets are a unique laboratory where chemical tagging can be performed to measure 
with high accuracy the elemental abundances of both stellar components, with the aim to investigate the 
formation of planets and their subsequent evolution. Here, we present a high-precision differential abundance analysis 
of the XO-2 wide stellar binary based on high resolution HARPS-N@TNG spectra. Both components are very similar 
K-dwarfs and host planets. Since they formed presumably within the same molecular cloud, we expect they should possess the 
same initial elemental abundances. We investigate if the presence of planets can cause some chemical imprints in the 
stellar atmospheric abundances. We measure abundances of 25 elements for both stars with a range of condensation 
temperature $T_{\rm C}=40-1741$ K, achieving typical precisions of $\sim 0.07$\,dex. The North component shows 
abundances in all elements higher by $+0.067 \pm 0.032$ dex on average, 
with a mean difference of +0.078 dex for elements with $T_{\rm C} > 800$\,K. The significance of the XO-2N abundance 
difference relative to XO-2S is at the $2\sigma$ level for almost all elements. We discuss the possibility that this result 
could be interpreted as the signature of the ingestion of material by XO-2N or depletion in XO-2S due to locking of 
heavy elements by the planetary companions. We estimate a mass of several tens of $M_{\oplus}$ 
in heavy elements. The difference in abundances between XO-2N and XO-2S shows a positive correlation with the 
condensation temperatures of the elements, with a slope of $(4.7 \pm 0.9) \times 10^{-5}$ dex K$^{-1}$, which could mean 
that both components have not formed terrestrial planets, but that first experienced the accretion of rocky core interior 
to the subsequent giant planets.
}
   
	   
\titlerunning{The GAPS programme with HARPS-N at TNG. X.}
\authorrunning{K. Biazzo et al.}
\maketitle

\section{Introduction}
\label{sec:intro}
A well-established direct dependence exists between the occurrence rate of giant planets and the metal content of their main-sequence 
hosts (\citealt{gonzalez1997, butleretal2000, lawsetal2003, santosetal2004, fischervalenti2005, udrysantos2007, sozzettietal2009, mortieretal2012}). 
Two main scenarios have been invoked to explain the observational evidence. On the one hand, the enhanced likelihood of forming giants planet in 
higher-metallicity disks can be seen as a direct consequence of the core-accretion scenario (\citealt{pollacketal1996, idalin2004, alibertetal2005}), 
because the expected larger reservoirs of dust grains in more metal-rich environments (e.g., \citealt{peifall1995, 
draineetal2007}) will favor the build-up of cores that will later accrete gas. In this case, the observed giant planet - metallicity connection 
has a primordial origin. On the other hand, it is possible that the outcome of the planet formation process might directly affect 
the chemical composition of the host stars. Two different versions of this second scenario have been proposed. In the 
first one, disk-planet or planet-planet interactions may cause the infall of metal-rich 
("rocky") material onto the star (\citealt{gonzalez1997}). In this case, the presence of planets is expected to positively correlate with 
metallicity. A similar effect is thought to be in action in the Solar System (\citealt{murrayetal2001}), where however the total mass of rocky 
material accreted by the Sun after the shrinking of its convective zone ($\sim 
0.4$~M$_\oplus$) is possibly too low for causing an appreciable variation of its chemical 
composition. Alternatively, the formation of planetary cores may prevent rocky planetary material to be accreted by the star (\citealt{melendezetal2009}). 
In this case, the presence of planets should negatively correlate with metallicity, in the sense that depletion of refractory elements relative 
to volatiles should be observed. Some evidence that this 
might be the case for the Sun has been presented by \cite{melendezetal2009}. \cite{chambers2010} 
estimated that the deficiency of refractory elements found by these authors would be canceled 
if $\sim 4~M_\oplus$ of rocky material would be added to the present-day solar convective envelope. 
\cite{melendezetal2009} also concluded that solar twins without close-in giant planets chemically resemble 
the Sun, suggesting that the presence of such planets might prevent the formation 
of Earth-size planets. Of course, it is possible that all these (and other) mechanisms are at work, rendering the discernment of 
their relative roles a rather complex undertaking (see, e.g., \citealt{gonzalezhernandezetal2013, adibekyanetal2014, maldonadoetal2015}).

While accretion of planetary material may cause huge effects in white dwarfs (e.g., 
\citealt{zuckermanetal2003}), variations of surface abundances are not expected to be large in 
FGK main sequence stars due to the dilution of the material within the outer convective 
envelope (\citealt{murrayetal2001}). In general, the most prominent effect should be an over- 
or under-abundance of elements as a function of the grains condensation temperature ($T_{\rm C}$), which 
is different for different elements (\citealt{smithetal2001}). Various authors have 
tried to show that such anomalies may be present in stars with planets (see, e.g., \citealt{smithetal2001, 
lawsgonzalez2001, melendezetal2009, ramirezetal2010, maldonadoetal2015}). Revealing these trends requires the 
development of rather ingenious techniques of very accurate differential abundance analysis (e.g., \citealt{ramirezetal2014}). 
In spite of these efforts, the results, if any, are still quite elusive. 
Obtaining accurate differential abundances is easier in the case of stars members of clusters (e.g., 
\citealt{yongetal2013}) and binary systems (\citealt{grattonetal2001, lawsgonzalez2001, 
desideraetal2004, desideraetal2006, ramirezetal2011, liuetal2014}), because here many observational 
uncertainties related to distances and reddening can be considered common-mode effects. In addition, the most accurate 
results are obtained comparing stars that are very similar to each other (\citealt{grattonetal2001, 
desideraetal2004, liuetal2014}). At the same time, it is also crucial that such stellar samples be accurately scrutinized for 
the presence of planets around them.

XO-2 is a wide visual binary (separation $\sim 31$~arcsec, corresponding to a projected distance of $\sim 4600$ AU; 
\citealt{burkeetal2007}) composed of two unevolved stars of very similar masses ($\sim 0.97 M_{\odot}$; \citealt{damassoetal2015}), 
and with the same proper motions (LSPM J0748+5013N and LSPM J0748+5013S in \citealt{lepineshara2005}). 
Both stars have no detectable amounts of lithium, supporting the old age of the system 
(\citealt{desideraetal2014, damassoetal2015}\footnote{In \cite{damassoetal2015}, we measured the XO-2 rotation periods, 
finding a value of 41.6 days in the case of XO-2N and a value between 26 and 34.5 days for XO-2S, combined with slight decreases in 
the activity levels of both stars at the end of the 2013-2014 season. Analyzing new APACHE photometric data taken from October 
2014 to April 2015, both these findings are not strengthened, probably as a result of the still decreased magnetic activity.}). 
A transiting giant planet was discovered around XO-2N by \cite{burkeetal2007}, with a period $P=2.616$~d, a mass 
$M_{\rm p}=0.597~M_{\rm J}$, and a semi-major axis 
$a=0.037$ AU. Recently, XO-2S was also shown to host a planetary system composed of at least two planets ($M_{\rm p}~\sin{i}=0.259~M_{\rm J}$, 
$P=18.157$~d, $a=0.134$ AU and $M_{\rm p}~\sin{i}=1.370~M_{\rm J}$, $P=120.80$~d, $a=0.476$ AU; \citealt{desideraetal2014}, 
hereafter Paper~I). In a subsequent characterization work on the XO-2 system, the existence was established of a long-term curvature 
in the radial velocity (RV) measurements of XO-2N, suggesting the possible presence of a second planet in a much wider orbit with $P\ge 17$~yr 
(\citealt{damassoetal2015}, hereafter Paper~II). The reason for the difference in the architectures of the planetary systems 
around the two stars is presently unclear. 

An accurate differential analysis of the iron content in the two components of XO-2 revealed a 
small but clearly significant difference of $0.054\pm 0.013$ dex, XO-2N being more iron-rich than XO-2S 
(Paper~II). This is a rare occurrence among binary systems, as in most cases the two components have 
very similar metal abundances (\citealt{desideraetal2004, desideraetal2006}). This result seems to be in contrast 
with the work by \cite{teskeetal2013}, who found the two stars similar in physical properties and in iron 
and nickel abundances, with some evidence of enrichment in C and O abundance of XO-2N compared to XO-2S. 
Very recently, \cite{teskeetal2015} explored the effects of changing the relative stellar temperature and 
gravity on the elemental abundances. Using three sets of parameters, they found that some refractory elements 
(Fe, Si, and Ni) are most probably enhanced in XO-2N regardless of the chosen parameters\footnote{After the present paper was 
submitted, an independent work by \cite{ramirezetal2015} based on different spectroscopic data and methodologies was published, 
with good agreement and consistent results with our analysis.}. Therefore, in the XO-2 system, similar to other binary systems 
hosting planets, the results are still ambiguous. In fact, there is still no consensus on the differences between 16\,Cyg\,A and B 
(\citealt{lawsgonzalez2001, takeda2005, schuleretal2011a, ramirezetal2011, metcalfeetal2012, tuccimaiaetal2014}, and references therein), 
while no significant abundance differences between components in the single-host systems HAT-P-1 (\citealt{liuetal2014}), HD\,132563 
(\citealt{desideraetal2011}), HD\,106515  (\citealt{desideraetal2012}) and the dual-host HD\,20782/1 system 
(\citealt{macketal2014}\footnote{HD\,20781 should host two Neptune-mass planets, as reported by Mayor et al. (2011), but the discovery 
was not confirmed/published.}) were found.

The XO-2 binary system is of particular interest. Being composed of two bright, near-twin stars with well-characterized planetary systems around them, 
it is one of only four known dual-planet-hosting binaries, together with HD\,20782/1 (\citealt{jonesetal2006}), Kepler\,132 (\citealt{roweetal2014}), and WASP\,94 
(\citealt{neveuvanmalleetal2014}). As we have mentioned above, planet-hosting binaries provide useful laboratories to try to decouple the effects of the stellar system 
initial abundances due to the star forming cloud from the possible effects of the planetary system on the star 
chemical composition. It is therefore important to perform 
a more extensive abundance analysis of the XO-2 system, not only to confirm the abundance difference we found for iron between two components, 
but also to establish if there is any correlation of the abundance differences with the condensation temperatures or other quantities. 

As a follow-on of Papers I and II and within the context of the GAPS\footnote{http://www.oact.inaf.it/exoit/EXO-IT/Projects/Entries/2011/12/27\_GAPS.html} 
(Global Architecture of Planetary Systems; \citealt{covinoetal2013}) programme, we present an accurate differential 
abundance analysis of 25 elements for the two components of the XO-2 system (XO-2N and XO-2S). The use 
of the procedure described in Paper II allows us to derive the differences in elemental 
abundances (and stellar parameters) with very high accuracy and unveil small differences among 
components, removing several sources of systematic errors. 

The outline of this paper is as follows. We first briefly present in Sect.\,\ref{sec:obs} 
the spectroscopic dataset. In Sect.\,\ref{sec:diff_abun_anal}, we describe the differential analysis of the elemental abundances for 25 species. 
We then discuss the behavior of the elemental abundance differences with the condensation temperature (Sect.\,\ref{sec:condensation}) 
and the origin of these differences (Sect.\,\ref{sec:discussion}). In Sect.\,\ref{sec:conclusions} our conclusions are presented. 

\section{Observations}
\label{sec:obs}
We observed both XO-2 components with the high resolution HARPS-N@TNG ($R \sim 115\,000$, $\lambda \sim 3\,900-6\,900$\,\AA; 
\citealt{cosentinoetal2012}) spectrograph between November 20, 2012 and October 4, 2014. Solar spectra were also obtained 
through observations of the asteroid Vesta. The spectra reduction was obtained using the 2013 November version of the HARPS-N 
instrument data reduction software (DRS) pipeline. A detailed description of the observations and data reduction is reported in 
Paper II.

As done in Paper II, we measured the elemental abundances co-adding the spectra 
of both components and Vesta, properly shifted by the corresponding radial velocity, to produce merged spectra with 
signal-to-noise ratios ($S/N$) around 300 (for the targets) and 400 (for the asteroid) per pixel at $\lambda \sim$6000 \AA. 

\section{Spectral analysis}
\label{sec:diff_abun_anal}
\subsection{Differential elemental abundances}
Elemental abundances\footnote{Throughout the paper, the abundance of the X element is given in $\log \frac{\epsilon{\rm (X)}}{\epsilon{\rm (H)}} + 12$.} 
were measured using the differential spectral analysis described in Paper II, which is based on the prescriptions 
widely described in \cite{grattonetal2001} and \cite{desideraetal2004,desideraetal2006}. In particular, this method gives very accurate 
results when the binary components are very similar in stellar parameters, as in the case of the XO-2 system. We thus considered 
an analysis based on equivalent widths (EWs), where a strict line-by-line method was applied, for both stars and the solar spectrum. 
Then, XO-2S was analyzed differentially with respect to the N component, which was used as a reference star. This allowed us to obtain 
the differences in stellar parameters and elemental abundances between XO-2N and XO-2S. This approach minimizes errors due to uncertainties 
in measurements of EWs, atomic parameters, solar parameters, and model atmospheres. Moreover, being the two components of this binary system 
very similar to each other, this reduces the possibility of systematic errors (\citealt{desideraetal2006}).

We considered as stellar parameters (effective temperature $T_{\rm eff}$, surface gravity $\log g$, microturbulence $\xi$) and 
iron abundances ([Fe/H]), together with their uncertainties, those reported in our Paper II, where \cite{kurucz1993} grids of 
plane-parallel model atmospheres were used. We refer to that work and \cite{biazzoetal2012} for a detailed description of the method 
to measure abundances (and their uncertainties). Once stellar parameters and iron abundance were set, we computed elemental abundances of several refractories, 
siderophiles, silicates, and volatiles elements (according to their cosmochemical character; see, e.g., Fig. 4 of \citealt{andersgrevesse1989}) 
using the MOOG code (\citealt{sneden1973}, version 2014) and the {\it abfind} and {\it blends} drivers for the treatment of the 
lines without and with hyperfine structure (HFS), respectively. 
We therefore obtained the abundance of 25 elements: C, N, O, Na, Mg, Al, Si, S, Ca, Sc, Ti, V, Cr, Mn, Fe, Co, Ni, Cu, Zn, Y, Zr, Ba, 
La, Nd, and Eu. For three elements (Ti, Cr, and Fe) we measured two ionization states, while for the other elements only one 
species (first or second ionization state) was measured. As done in Paper II for the iron lines, here the EWs of each spectral 
line were measured on the one-dimensional spectra interactively using the {\it splot} task in IRAF\footnote{IRAF is distributed by the
National Optical  Astronomy Observatory, which is operated by the Association of the Universities for Research in Astronomy, Inc. (AURA) 
under cooperative agreement with the National Science Foundation.}. The location of the local continuum 
was carefully selected tracing as much as possible the same position for each spectral line of both binary system components and 
the asteroid Vesta. This was done with the aim to minimize the error in the selection of the continuum. We also excluded features 
affected by telluric absorption.

We compiled a long list of spectral lines, starting from the work by \cite{biazzoetal2012}. In particular, our line list comprises 
\ion{Na}{i}, \ion{Mg}{i}, \ion{Al}{i}, \ion{Si}{i}, \ion{Ca}{i}, \ion{Ti}{i}, \ion{Ti}{ii}, \ion{Cr}{i}, \ion{Cr}{ii}, \ion{Fe}{i}, 
\ion{Fe}{ii}, \ion{Ni}{i}, and \ion{Zn}{i}, complemented with additional lines and atomic parameters for \ion{Na}{i}, \ion{Al}{i}, 
\ion{Si}{i}, \ion{Ti}{i}, \ion{Ti}{ii}, \ion{Cr}{i}, \ion{Ni}{i}, and \ion{Zn}{i} taken from \cite{schuleretal2011b} and \cite{sozzettietal2006} 
to increase the statistics. In seven cases (\ion{C}{i}, \ion{S}{i}, \ion{Sc}{ii}, \ion{V}{i}, \ion{Mn}{i}, \ion{Co}{i}, and \ion{Cu}{i}), 
we considered the line lists by \cite{kurucz1993}, \cite{schuleretal2011b}, \cite{johnsonetal2006}, and \cite{scottetal2015b}, where 
the HFS by \cite{johnsonetal2006} and \cite{kurucz1993} was adopted for Sc, V, Mn, Cu, and Co. Solar isotopic ratios by 
\cite{andersgrevesse1989} were considered for Cu (i.e. 69.17\% for $^{63}$Cu and 30.83\% for $^{65}$Cu). For the $s$-process elements 
\ion{Y}{ii}, \ion{Zr}{ii} (first peak) and \ion{Ba}{ii}, \ion{La}{ii} (second peak) and the $r$-process elements \ion{Nd}{ii} and 
\ion{Eu}{ii} we considered the line lists by \cite{johnsonetal2006}, \cite{ljungetal2006}, \cite{prochaskaetal2000}, 
\cite{lawleretal2001b}, and \cite{denhartogetal2003} where the HFS by \cite{gallagheretal2010} and \cite{lawleretal2001a} was adopted 
for Ba and La, respectively. Solar isotopic ratios by \cite{andersgrevesse1989} were considered for Ba (i.e. 2.417\% for $^{134}$Ba, 
7.854\% for $^{136}$Ba, 71.70\% for $^{138}$Ba, and 6.592\% for $^{135}$Ba, 11.23\% for $^{137}$Ba) and Eu (i.e. 47.8\% for $^{151}$Eu 
and 52.2\% for $^{153}$Eu). 

\begin{figure}[t!]
\begin{center}
\includegraphics[width=8.5cm]{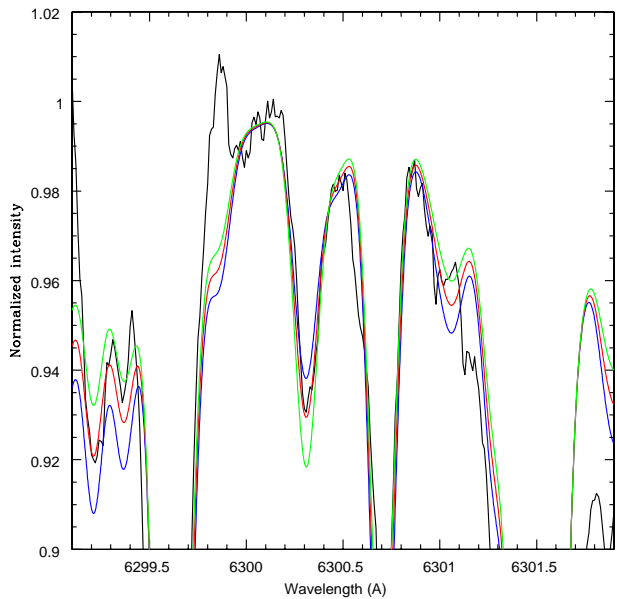}
\caption{Comparison between observed (black line) and synthetic spectra for XO-2S in the region around the [\ion{O}{i}] line at 6300.3~\AA. 
The synthetic spectra have been computed for $\log{n{\rm (O)}}=9.15$~(blue line), 9.20 (red line), and 9.25 (green line), respectively. 
Note that several CN lines are also present in this spectral region; due to coupling between C and O, the strength of these
features is anti-correlated with that of O.}
\label{fig:o} 
\end{center}
\end{figure}

\begin{figure}[t!]
\begin{center}
\includegraphics[width=8.5cm]{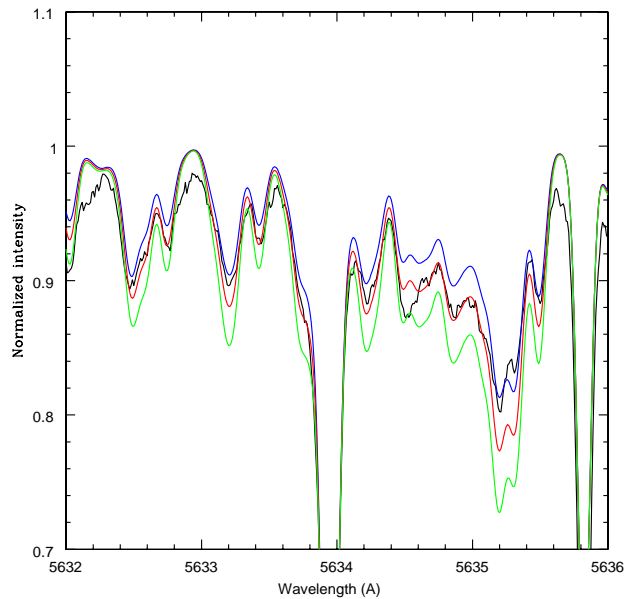}
\caption{Comparison between observed (black line) and synthetic spectra for XO-2S in the region around the bandhead of C$_2$\
at 5635~\AA. The synthetic spectra have been computed for $\log{n{\rm (C)}}=8.83$~(blue line), 8.88 (red line), and 8.93 (green line), 
respectively. }
\label{fig:c2} 
\end{center}
\end{figure}

CNO abundances were measured as follows. First, O was measured by using the {\it synth} driver within the MOOG code (\citealt{sneden1973}, 
version 2014) fitting synthetic spectra to the zone around the [\ion{O}{i}] line at 6300.3\,\AA\,(see Fig.~\ref{fig:o}). Then adopting the 
estimated O abundance, C was measured from features due to the C$_2$ Swan system at 5150 ((0,0) band) and 5635\,\AA\,((0,1) band: see Fig.~\ref{fig:c2}). 
Finally, N was measured from several CN features at $\sim$4195 and $\sim$4215\,\AA, adopting the measured C and O abundances.
The process was repeated for both the stars and for the Vesta spectrum until convergence of the derived values. Note that the procedure
was strictly differential, using the same features in each of the stars. Line lists were built using the latest version of the VALD 
database\footnote{http://vald.astro.uu.se/} for the atomic lines, while the line lists for CN are from  \citet{sneden2014}  and C$_2$ are from
\citet{rametal2014} and \citet{brookeetal2013}. Carbon abundances were also derived from atomic \ion{C}{i}. Since C abundance was derived with both EWs and 
spectral synthesis, as final abundance we considered the mean value coming from these two methods.

Table \ref{tab:solar_abundances} lists the solar elemental abundances we obtained using spectra of the Vesta asteroid and the 
following solar parameters: $T_{\rm eff \odot}=5770$ K, $\log g_\odot=4.44$, and $\xi_\odot=1.10$ km/s (see, e.g., \citealt{biazzoetal2012}). In the same table, the 
most recent determinations of solar abundances by \cite{asplundetal2009}, \cite{scottetal2015a}, \cite{scottetal2015b}, and \cite{grevesseetal2015} are 
also given. We stress that the latter values were obtained using 3D+NLTE models. Our determinations are therefore in good agreement 
with these recent literature values.

\begin{figure}[t!]
\begin{center}
 \begin{tabular}{c}
\hspace{+.2cm}
\includegraphics[width=8.5cm]{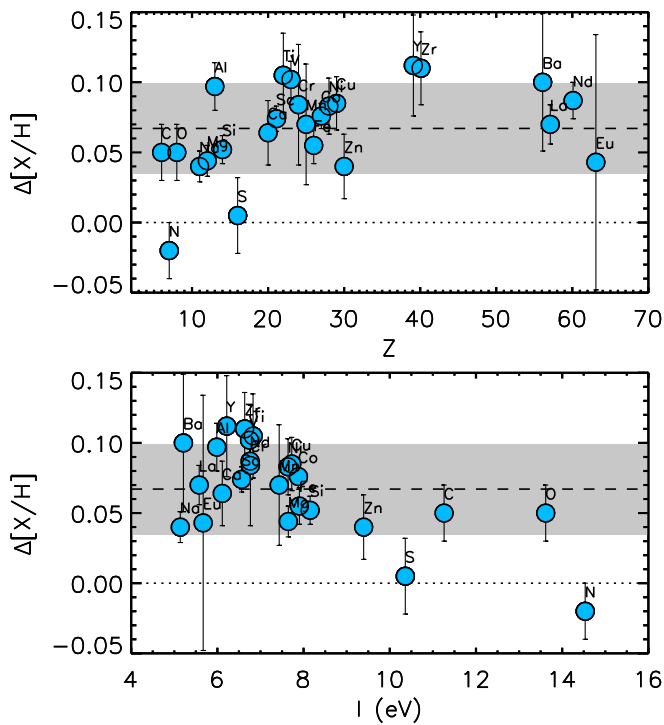}
 \end{tabular}
\vspace{-.6cm}
\caption{Elemental abundance difference between XO-2N and XO-2S as a function of the atomic number Z ({\it upper panel}) and 
ionization potential $I$ ({\it lower panel}). In both panels, dotted line refers to null difference, dashed line represents 
the mean value of the difference, and the filled grey area is the $\pm 1 \sigma$ level.}
\label{fig:XO2system_dXH_ZI} 
 \end{center}
\end{figure}

\begin{figure}[t!]
\begin{center}
 \begin{tabular}{c}
\hspace{+.2cm}
\includegraphics[width=8.5cm]{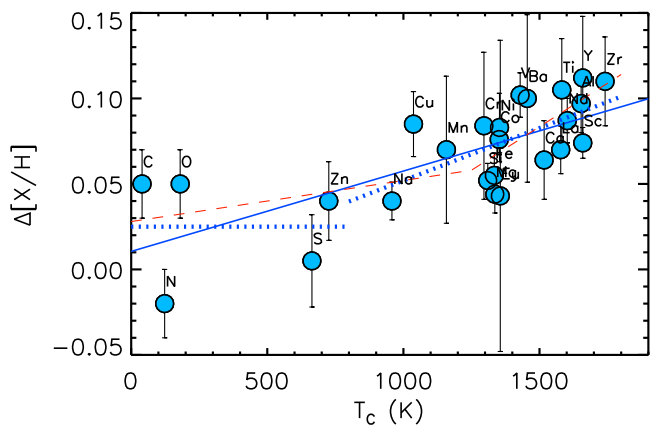}
\vspace{.4cm}
\end{tabular}
\caption{Differential elemental abundances of XO-2N\,$-$\,XO-2S versus condensation temperatures. Solid blue line is the unweighted linear 
least-square fit to our data. The horizontal dotted blue line represents the average of the volatiles, while the other dotted blue line is the trend 
of the refractory elements. The dashed red line is the mean trend obtained by \cite{melendezetal2009} for eleven solar twins 
with respect to the Sun, after a vertical shift was applied to match the refractory elements.}
\label{fig:XO2system_dXH_Tc} 
 \end{center}
\end{figure}

In Table \ref{tab:elemental_abundances} we present the final elemental abundances (relative to the solar ones) and the 
differential abundances (N$-$S) of all elements, together with their respective errors. The estimate of the uncertainties on 
each elemental abundance and on differential [X/H] was 
done as widely described in Paper II (see also references therein). 
Table \ref{tab:errors} lists the mean internal errors in abundances determination due to uncertainties in stellar parameters for both components, 
where the stellar parameters and iron abundance are those obtained and reported in Paper II. In the end, the differences in 
parameters (in the sense of N minus S component) are: 
$\Delta T_{\rm eff}= -35 \pm 8$ K, $\Delta \log g = +0.010 \pm 0.020$ dex, $\Delta \xi= -0.07 \pm 0.07$ km s$^{-1}$, 
and [Fe/H]$=+0.055\pm0.013$ dex. Our analysis demonstrates that: $i)$ the S component is slightly hotter than the N component, 
as reported in Paper II; $ii)$ besides the enhancement in iron found in our Paper II, the XO-2 binary system is more rich than 
the Sun in almost all elements (see Fig.~\ref{fig:XO2system_dXH_ZI}, where the differential abundances of the N minus S component are 
displayed as a function of the atomic number Z and the ionization potential $I$); $iii)$ XO-2N is enhanced in all refractory elements 
and less enhanced in average in volatiles (see Fig.~\ref{fig:XO2system_dXH_Tc}, where the differential abundances are displayed as a function 
of the condensation temperatures. Hints of a similar behavior were reported by \cite{teskeetal2015} for Fe, Si, and Ni.

All differential elemental abundances have errors in the range $\sim 0.01-0.05$ dex (with the exception of the europium), which 
further demonstrates the advantages of a strictly differential analysis. The mean abundance difference of all elements between XO-2N and XO-2S is 
$\Delta$[X/H]$ = +0.067 \pm 0.032$ dex, with a mean difference of $0.025$ dex for the volatile elements ($T_{\rm c}<800$\,K) and of $\sim 0.078$\,dex 
for the elements with $T_{\rm C}> 800$\,K. For the neutron-capture elements (e.g., \ion{Y}{ii}, \ion{Zr}{ii}, \ion{Ba}{ii}, \ion{La}{ii}, \ion{Nd}{ii}, 
\ion{Eu}{ii}) the mean difference is 0.087\,dex. No elemental abundance of XO-2N differs from that of XO-2S by more than $\sim 0.11$ dex, and all 
but three species (namely, N, S, and Eu) have differences detected at $\gtsim 2 \sigma$ level.

\begin{table}[h]
\caption{Comparison between our measured solar abundances and standard values from the literature.} 
\label{tab:solar_abundances}
\begin{center}
\begin{tabular}{lccc}
\hline
\hline
Species & $\log n_{\rm This\,work}$ & $\log n_{\rm Literature}$ & Reference\\ 
\hline
C            & $8.50  \pm 0.03$  & $8.43 \pm 0.05$ & (1) \\
N            & $7.93  \pm 0.04$  & $7.83 \pm 0.05$ & (1) \\
O            & $8.93  \pm 0.05$  & $8.69 \pm 0.05$ & (1) \\
Na           & $6.305 \pm 0.007$ & $6.21 \pm 0.04$ & (2) \\
Mg           & $7.590 \pm 0.044$ & $7.59 \pm 0.04$ & (2) \\
Al           & $6.490 \pm 0.115$ & $6.43 \pm 0.04$ & (2) \\
Si           & $7.558 \pm 0.052$ & $7.51 \pm 0.03$ & (2) \\
S            & $7.210 \pm 0.014$ & $7.12 \pm 0.03$ & (2) \\
Ca           & $6.347 \pm 0.035$ & $6.32 \pm 0.03$ & (2) \\
Sc           & $3.135 \pm 0.049$ & $3.16 \pm 0.04$ & (2) \\
\ion{Ti}{i}  & $4.960 \pm 0.043$ & $4.93 \pm 0.04$ & (3) \\
\ion{Ti}{ii} & $4.982 \pm 0.048$ &                 &     \\
V            & $3.856 \pm 0.039$ & $3.89 \pm 0.08$ & (3) \\
\ion{Cr}{i}  & $5.662 \pm 0.033$ & $5.62 \pm 0.04$ & (3) \\
\ion{Cr}{ii} & $5.655 \pm 0.049$ &		   &	 \\
Mn           & $5.407 \pm 0.064$ & $5.42 \pm 0.04$ & (3) \\
\ion{Fe}{i}  & $7.530 \pm 0.049$ & $7.47 \pm 0.04$ & (3) \\
\ion{Fe}{ii} & $7.535 \pm 0.053$ &		   &	 \\
Co           & $4.894 \pm 0.074$ & $4.93 \pm 0.05$ & (3) \\
Ni           & $6.279 \pm 0.048$ & $6.20 \pm 0.04$ & (3) \\
Cu           & $4.213 \pm 0.074$ & $4.18 \pm 0.05$ & (4) \\
Zn           & $4.553 \pm 0.025$ & $4.56 \pm 0.05$ & (4) \\
Y            & $2.165 \pm 0.051$ & $2.21 \pm 0.05$ & (4) \\
Zr           & $2.600 \pm 0.031$ & $2.59 \pm 0.04$ & (4) \\
Ba           & $2.235 \pm 0.007$ & $2.25 \pm 0.07$ & (4) \\
La           & $1.105 \pm 0.007$ & $1.11 \pm 0.04$ & (4) \\
Nd           & $1.460 \pm 0.044$ & $1.42 \pm 0.04$ & (4) \\
Eu           & $0.543 \pm 0.042$ & $0.52 \pm 0.04$ & (4) \\
\hline
\end{tabular}
\end{center}
Notes: (1) \cite{asplundetal2009}; (2) \cite{scottetal2015b}; (3) \cite{scottetal2015a}; (4) \cite{grevesseetal2015}.
\end{table}

\setlength{\tabcolsep}{4.2pt}
\begin{table}
\caption{Differential elemental abundances for the XO-2 system.} 
\label{tab:elemental_abundances}
\begin{center}
\small
\begin{tabular}{lcccc}
\hline
\hline
Element  & $T_{\rm C}^a$ & XO-2N$^b$ &  XO-2S$^b$ & $\Delta[$X/H$]^c$  \\ 
\hline
$[$C/H$]$       & 40   & $0.41 \pm 0.04$   & $0.36 \pm 0.05$   & $ 0.050 \pm 0.02$ \\
$[$N/H$]$       & 123  & $0.47 \pm 0.05$   & $0.49 \pm 0.05$   & $-0.020 \pm 0.02$ \\
$[$O/H$]$       & 180  & $0.32 \pm 0.03$   & $0.27 \pm 0.03$   & $ 0.050 \pm 0.02$ \\
$[$Na/H$]$      & 958  & $0.485 \pm 0.043$ & $0.445 \pm 0.043$ & $ 0.040 \pm 0.011$ \\
$[$Mg/H$]$      & 1336 & $0.377 \pm 0.051$ & $0.333 \pm 0.048$ & $ 0.044 \pm 0.011$ \\
$[$Al/H$]$      & 1653 & $0.440 \pm 0.139$ & $0.343 \pm 0.141$ & $ 0.097 \pm 0.017$ \\
$[$Si/H$]$      & 1310 & $0.471 \pm 0.082$ & $0.419 \pm 0.082$ & $ 0.052 \pm 0.010$ \\
$[$S/H$]$       & 664  & $0.530 \pm 0.020$ & $0.525 \pm 0.025$ & $ 0.005 \pm 0.027$ \\
$[$Ca/H$]$      & 1517 & $0.365 \pm 0.075$ & $0.301 \pm 0.065$ & $ 0.064 \pm 0.023$ \\
$[$Sc/H$]$      & 1659 & $0.462 \pm 0.067$ & $0.388 \pm 0.060$ & $ 0.074 \pm 0.009$ \\
$[$Ti/H$]^\ast$ & 1582 & $0.456 \pm 0.096$ & $0.351 \pm 0.091$ & $ 0.105 \pm 0.030$ \\
$[$V/H$]$       & 1429 & $0.431 \pm 0.060$ & $0.329 \pm 0.062$ & $ 0.102 \pm 0.013$ \\
$[$Cr/H$]^\ast$ & 1296 & $0.423 \pm 0.059$ & $0.339 \pm 0.054$ & $ 0.084 \pm 0.043$ \\
$[$Mn/H$]$      & 1158 & $0.576 \pm 0.065$ & $0.506 \pm 0.074$ & $ 0.070 \pm 0.043$ \\
$[$Fe/H$]^\ast$ & 1334 & $0.370 \pm 0.072$ & $0.315 \pm 0.078$ & $ 0.055 \pm 0.013$ \\
$[$Co/H$]$      & 1352 & $0.524 \pm 0.127$ & $0.448 \pm 0.126$ & $ 0.076 \pm 0.009$ \\
$[$Ni/H$]$      & 1353 & $0.538 \pm 0.094$ & $0.455 \pm 0.096$ & $ 0.083 \pm 0.020$ \\
$[$Cu/H$]$      & 1037 & $0.627 \pm 0.109$ & $0.542 \pm 0.107$ & $ 0.085 \pm 0.019$ \\
$[$Zn/H$]$      & 726  & $0.467 \pm 0.044$ & $0.427 \pm 0.059$ & $ 0.040 \pm 0.023$ \\
$[$Y/H$]$       & 1659 & $0.472 \pm 0.134$ & $0.360 \pm 0.110$ & $ 0.112 \pm 0.036$ \\
$[$Zr/H$]$      & 1741 & $0.460 \pm 0.084$ & $0.350 \pm 0.084$ & $ 0.110 \pm 0.026$ \\
$[$Ba/H$]$      & 1455 & $0.410 \pm 0.022$ & $0.310 \pm 0.049$ & $ 0.100 \pm 0.049$ \\
$[$La/H$]$      & 1578 & $0.530 \pm 0.036$ & $0.460 \pm 0.022$ & $ 0.070 \pm 0.014$ \\
$[$Nd/H$]$      & 1602 & $0.547 \pm 0.151$ & $0.460 \pm 0.147$ & $ 0.087 \pm 0.013$ \\
$[$Eu/H$]$      & 1356 & $0.520 \pm 0.108$ & $0.477 \pm 0.104$ & $ 0.043 \pm 0.091$ \\
\hline
\end{tabular}
\end{center}
Notes: $^a$ As condensation temperatures we considered the 50\% $T_{\rm C}$ values derived by \cite{lodders2003} 
at a total pressure of $10^{-4}$ bar. $^b$ Abundances relative to the Sun. $^c$ Abundances of XO-2N relative to XO-2S. 
Errors refer to the differential elemental abundances (see text). 
$^\ast$ Ti, Cr, and Fe abundances refer to the values obtained with the first ionization state.
\end{table}
\normalsize

\subsection{Stellar population of the XO-2 components}

Elemental abundances, together with Galactic kinematics, are often used to identify a star as belonging to a specific 
stellar population. This is because the thin disk and the thick disk within the Milky Way are discrete populations showing distinct 
chemical distributions in several elements, and in particular in $\alpha$-elements. In fact, the Galactic thin and thick disks 
appear to overlap significantly 
around $-0.7 \ltsim$[Fe/H]$\ltsim 0.0$ dex when the iron abundance is used as a reference, while they are separated in 
[$\alpha$/Fe] (see, e.g., \citealt{bensbyetal2003, misheninaetal2004}). 

\begin{figure}[t!]
\begin{center}
 \begin{tabular}{c}
\hspace{.3cm}
\includegraphics[width=9cm]{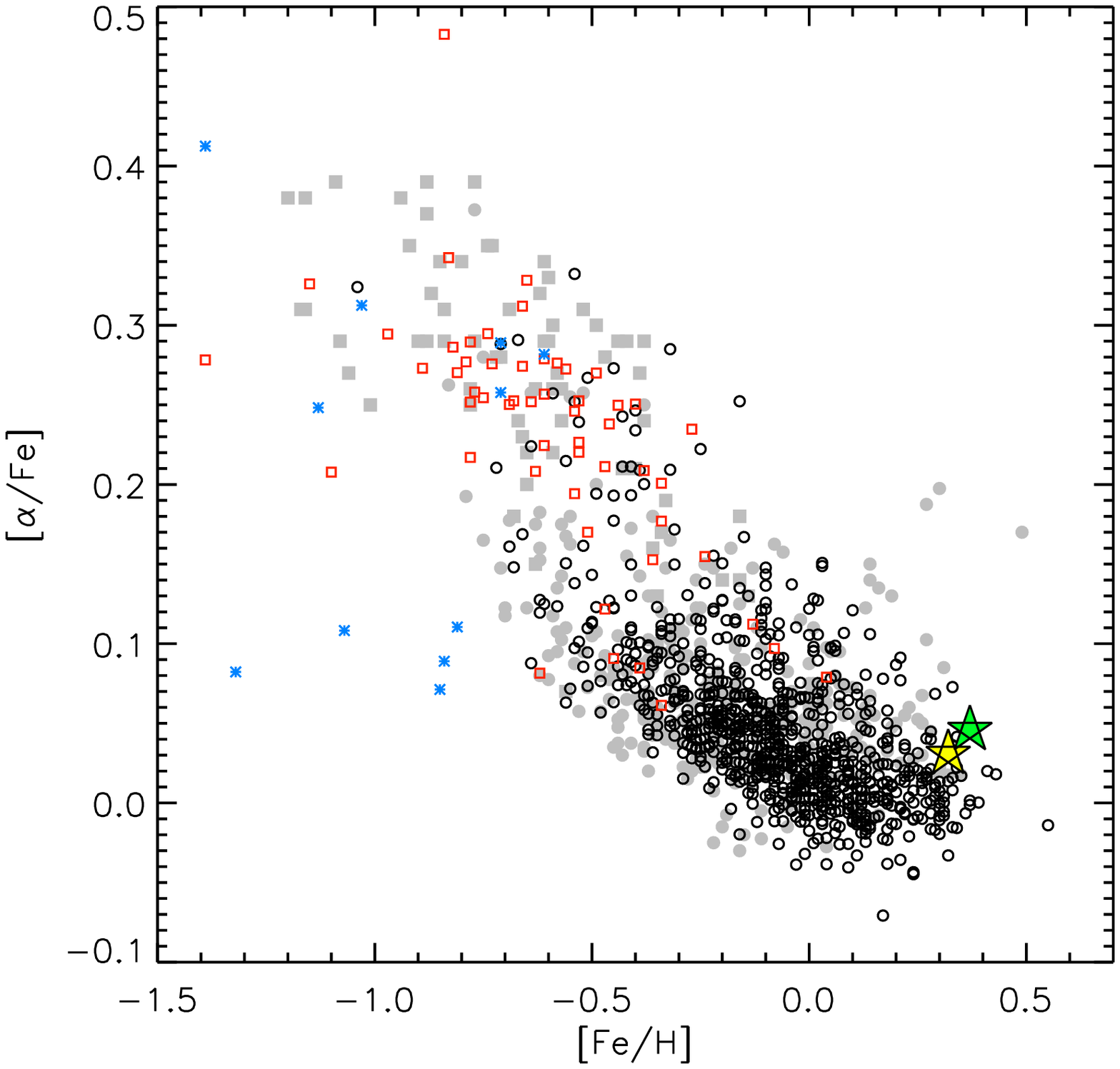}
 \end{tabular}
\caption{[$\alpha$/Fe] versus [Fe/H] for the \cite{adibekyanetal2012} sample, where open circles, open squares, and asterisks represent 
stars with $P_{\rm thin} \ge 0.90$, $P_{\rm thick} \ge 0.90$, and $P_{\rm halo} \ge 0.90$ (from \citealt{bensbyetal2003}), respectively. 
Filled circles and squares show the stars within the \cite{soubirangirard2005} sample with $P_{\rm thin} \ge 0.90$ and $P_{\rm thick} \ge 0.90$ 
(from \citealt{misheninaetal2004}), respectively. The positions of the two targets XO-2N and XO-2S are displayed with big yellow and green 
five-pointed star symbols, respectively.}
\label{fig:population_XO2} 
 \end{center}
\end{figure}

Here, we investigate to which population the XO-2 stellar system belongs to taking advantage of 431 stars within the 
catalogue by \cite{soubirangirard2005} and 906 FGK dwarf stars analyzed by \cite{adibekyanetal2012} with thin-disk, thick-disk, 
and halo membership probabilities ($P_{\rm thin}$, $P_{\rm thick}$, $P_{\rm halo}$) higher than 90\%. We use as abundance of 
$\alpha$-elements the definition [$\alpha$/Fe]=$\frac{1}{4}$([Mg/Fe]+[Si/Fe]+[Ca/Fe]+[Ti/Fe]), since magnesium, silicon, calcium, 
and titanium are the common elements analyzed by those authors and by us, and also because they show similar behavior in the 
chemical evolution within our Galaxy. Figure \ref{fig:population_XO2} displays the position of the XO-2N and XO-2S stars in 
the [$\alpha$/Fe] versus [Fe/H] diagram. Based on this sort of ``chemical indicator'', the stellar system studied in this work 
appears to belong to the Galactic thin disk population. This result gives support to previous findings based on kinematics 
(\citealt{burkeetal2007}) and analysis of the Galactic orbits (Paper II).

\section{Abundance differences versus condensation temperatures} 
\label{sec:condensation}

Possible evidences of planetesimal accretion can be investigated by searching for any dependence between elemental abundances 
and condensation temperatures. This is because any accretion event would occur very close to the star (i.e. in a 
high-temperature environment). Since refractory elements (e.g., $\alpha$- and Fe-peak elements) condense at high $T_{\rm C}$, 
they might be added in large quantities with respect to volatiles elements (e.g., C, N, O, Zn) with lower $T_{\rm C}$ 
(\citealt{sozzettietal2006}, and references therein).

Any trend of [X/H] with  $T_{\rm C}$ can be quantified in terms of a positive slope in a linear fit. The possibility of a trend of [X/H] 
with $T_{\rm C}$ (\citealt{sozzettietal2006, gonzalezhernandezetal2013, liuetal2014, macketal2014}, and references therein) and its 
dependence on stellar age, surface gravity, and mean galactocentric distance (\citealt{adibekyanetal2014}) has been investigated 
by several authors, and often no statistically convincing trend could be found.

The similarity in stellar parameters between the XO-2 binary components allowed us to derive differential elemental abundances with a 
high level of confidence. Figure \ref{fig:XO2system_dXH_Tc} shows the differential abundances of the XO-2N and XO-2S stars relative to 
each other versus $T_{\rm C}$ in the range 40-1741 K (see also Table~\ref{tab:elemental_abundances}). We remark the importance to 
have as much as possible an extended range of condensation temperatures to put constraints in such a trend. The slope of the unweighted linear 
least-square fit is positive, with a value of $(4.7 \pm 0.9) \times 10^{-5}$ dex K$^{-1}$. We calculated the Pearson's $r$ correlation coefficient by 
means of the IDL\footnote{IDL (Interactive Data Language) is a registered trademark of Exelis Visual Information Solutions.} procedure CORRELATE 
(\citealt{pressetal1986}). This test indicates that $\Delta$[X/H] and $T_{\rm C}$ are positively correlated, with $r \sim 0.72$. 
Considering only the elements with higher values of $T_{\rm C}$ (i.e. $> 800$ K), the slope and $r$ become $(6.1 \pm 2.1) \times 10^{-5}$ dex K$^{-1}$ 
and $\sim 0.57$, respectively. Moreover, using Monte Carlo simulations (\citealt{parkeloydfrance2014}), we found that 
the probability that uncorrelated random datasets can reproduce the observed arrangement of point is only $\sim 0.9\%$. This proves that the 
observed positive correlation between the differential elemental abundances of XO-2N related to XO-2S and the condensation 
temperatures is significant. 
A similar peculiar positive trend in the observed differential abundances with $T_{\rm C}$ has been found in solar twin stars known to 
have close-in giant planets (see Fig.~\ref{fig:XO2system_dXH_Tc}), while the majority of solar analogues without revealed giant planets in radial velocity monitoring 
show a negative trend, such as the Sun (\citealt{melendezetal2009}). This result would seem to be consistent with the expectation 
that the presence of close-in giant planets prevents the formation and/or survival of terrestrial planets (\citealt{idalin2004}). 

\subsection{Comparison with previous works}

In Paper II we have found that XO-2N is richer in Fe abundance than the S component. Here, we confirm this finding for many more 
elements (in particular, the refractories). This result seems to be consistent with the recent work by \cite{teskeetal2015}, 
who found that the N component is most probably enhanced in Si, Fe, and Ni when compared to XO-2S. In the case of 16\,Cyg A/B, no 
consensus was reached, as some authors 
find differences in the differential abundances (e.g., \citealt{ramirezetal2011, tuccimaiaetal2014}), and others claim the stars 
are chemically homogeneous (with the exception of Li and B). \cite{macketal2014} find non-zero differences for O, Al, V, Cr, Fe, 
and Co of HD20781/2. Again, \cite{liuetal2014} found $\Delta$[X/H] values consistent with zero in the HAT-P-1 G0+F8 binary. 
Recently, \cite{teskeetal2015} claim that perhaps we are seeing an effect in $\Delta$[X/H] trends due to $T_{\rm eff}$, 
moving from hotter stars with no differences to cooler stars with more significant differences. They also warn that we have 
to be cautious about this question, as it is hard at the present day to have an answer due to the different analysis methods and 
the host star nature of these systems (i.e. whether or not they host planets). Moreover, as pointed out by \cite{sozzettietal2006}, 
there is an intrinsic difficulty in determining with accuracy elemental abundances for a large set of elements in cool stars without 
the danger to introduce greater uncertainties in the results.

Similarly to the analysis of several binary systems (see, e.g., \citealt{ramirezetal2009, ramirezetal2011, schuleretal2011b, 
liuetal2014, macketal2014, tuccimaiaetal2014}), we find a positive trend with $T_{\rm C}$ for the elemental abundances. 
Our results are consistent with accretion events of H-depleted rocky material on both components. In single stars, 
an unambiguous explanation for these trends is difficult to reach, even in presence of slopes in the $\Delta$[X/H]$-T_C$ relation 
steeper than Galactic chemical evolution effects, as they may not indicate necessarily the presence or absence of planets. 
\cite{gonzalezhernandezetal2013} have found that after removing the Galactic chemical evolution effects in a sample of 61 single FG-type 
main-sequence objects, stars both with and without planets show similar mean abundance patterns. On the other hand, Galactic 
chemical evolution is not capable of producing specific element-by-element difference between components of a binary system, 
whereas the planetary accretion scenario might reproduce it naturally (see \citealt{macketal2014}).

\setlength{\tabcolsep}{2.pt}
\begin{table}  
\caption{Mean internal errors in abundance determinations due to given uncertainties ($\Delta T_{\rm eff}$, $\Delta \log g$, and $\Delta \xi$) 
in stellar parameters for XO-2N and XO-2S.}
\label{tab:errors}
\begin{center}
\small
\begin{tabular}{lccc}
\hline
\hline
XO-2N & $T_{\rm eff}=5290$ K & $\log g=4.43$ & $\xi=0.86$ km/s\\
\hline
$\Delta$   & $\Delta T_{\rm eff}=\pm 18$ K & $\Delta \log g=\pm 0.10$ & $\Delta \xi=\pm 0.06$ km/s\\
\hline
$[$C/H$]$             &  $0.02$    & $0.05$  & $0.02$	\\
$[$N/H$]$             &  $0.01$    & $0.05$  & $0.01$	\\
$[$O/H$]$             &  $0.00$    & $0.05$  & $0.02$	\\
$[$Na/H$]$            &  $0.013$   & $0.003$ & $0.010$  \\
$[$Mg/H$]$            &  $0.003$   & $0.002$ & $0.005$  \\
$[$Al/H$]$            &  $0.013$   & $0.013$ & $0.010$  \\
$[$Si/H$]$            &  $0.005$   & $0.001$ & $0.003$  \\
$[$S/H$]$             &  $0.010$   & $0.006$ & $0.006$  \\
$[$Ca/H$]$            &  $0.016$   & $0.002$ & $0.015$  \\
$[$Sc/H$]$            &  $0.002$   & $0.003$ & $0.005$  \\
$[$\ion{Ti}{i}/H$]$   &  $0.022$   & $0.002$ & $0.026$  \\
$[$\ion{Ti}{ii}/H$]$  &  $0.003$   & $0.002$ & $0.016$  \\
$[$V/H$]$             &  $0.024$   & $0.001$ & $0.008$  \\	 
$[$\ion{Cr}{i}/H$]$   &  $0.017$   & $0.002$ & $0.016$  \\
$[$\ion{Cr}{ii}/H$]$  &  $0.009$   & $0.003$ & $0.013$  \\
$[$Mn/H$]$            &  $0.017$   & $0.001$ & $0.029$  \\	 
$[$\ion{Fe}{i}/H$]$   &  $0.010$   & $0.001$ & $0.015$  \\
$[$\ion{Fe}{ii}/H$]$  &  $0.013$   & $0.004$ & $0.014$  \\
$[$Co/H$]$            &  $0.008$   & $0.002$ & $0.003$  \\	 
$[$Ni/H$]$            &  $0.003$   & $0.001$ & $0.018$  \\   
$[$Cu/H$]$            &  $0.006$   & $0.003$ & $0.017$  \\    
$[$Zn/H$]$            &  $0.005$   & $0.001$ & $0.019$  \\	  
$[$Y/H$]$             &  $0.003$   & $0.003$ & $0.030$  \\	  
$[$Zr/H$]$            &  $0.001$   & $0.003$ & $0.023$  \\	 
$[$Ba/H$]$            &  $0.006$   & $0.001$ & $0.043$  \\	 
$[$La/H$]$            &  $0.003$   & $0.006$ & $0.003$  \\	 
$[$Nd/H$]$            &  $0.004$   & $0.004$ & $0.007$  \\	 
$[$Eu/H$]$            &  $0.001$   & $0.006$ & $0.001$  \\	  
\hline	\\	
XO-2S & $T_{\rm eff}=5325$ K & $\log g=4.420$ & $\xi=0.93$ km/s\\
\hline
$\Delta$   & $\Delta T_{\rm eff}=\pm 37$ K & $\Delta \log g=\pm 0.094$ & $\Delta \xi=\pm 0.03$ km/s\\
\hline
$[$C/H$]$           & $0.04 $ & $0.05$  & $0.01$   \\
$[$N/H$]$           & $0.02 $ & $0.05$  & $0.01$ \\
$[$O/H$]$           & $0.00 $ & $0.05$  & $0.01$ \\
$[$Na/H$]$          & $0.023$ & $0.020$ & $0.003$  \\
$[$Mg/H$]$          & $0.013$ & $0.015$ & $0.002$  \\
$[$Al/H$]$          & $0.022$ & $0.007$ & $0.003$  \\
$[$Si/H$]$          & $0.011$ & $0.001$ & $0.004$    \\
$[$S/H$]$           & $0.020$ & $0.020$ & $0.001$   \\
$[$Ca/H$]$          & $0.030$ & $0.026$ & $0.009$ \\
$[$Sc/H$]$          & $0.006$ & $0.034$ & $0.004$ \\
$[$\ion{Ti}{i}/H$]$ & $0.043$ & $0.009$ & $0.013$ \\
$[$\ion{Ti}{ii}/H$]$& $0.008$ & $0.029$ & $0.010$  \\
$[$V/H$]$           & $0.046$ & $0.006$ & $0.003$  \\			   
$[$\ion{Cr}{i}/H$]$ & $0.034$ & $0.016$ & $0.015$   \\
$[$\ion{Cr}{ii}/H$]$& $0.020$ & $0.029$ & $0.005$\\
$[$Mn/H$]$          & $0.035$ & $0.015$ & $0.015$ \\		     
$[$\ion{Fe}{i}/H$]$ & $0.020$ & $0.011$ & $0.008$ \\
$[$\ion{Fe}{ii}/H$]$& $0.026$ & $0.035$ & $0.006$ \\
$[$Co/H$]$          & $0.017$ & $0.017$ & $0.003$  \\			   
$[$Ni/H$]$          & $0.006$ & $0.004$ & $0.009$  \\			
$[$Cu/H$]$          & $0.016$ & $0.006$ & $0.006$  \\			
$[$Zn/H$]$          & $0.012$ & $0.005$ & $0.009$ \\		     
$[$Y/H$]$           & $0.002$ & $0.028$ & $0.015$ \\		     
$[$Zr/H$]$          & $0.002$ & $0.037$ & $0.009$  \\		       
$[$Ba/H$]$          & $0.006$ & $0.015$ & $0.020$  \\		       
$[$La/H$]$          & $0.005$ & $0.040$ & $0.001$\\		  
$[$Nd/H$]$          & $0.005$ & $0.039$ & $0.003$   \\  	       
$[$Eu/H$]$          & $0.001$ & $0.040$ & $0.001$\\		     
\hline	\\	
\end{tabular}
\end{center}
\end{table}
\normalsize

\section{Origin of the elemental abundances difference}
\label{sec:discussion}

The two stars have distinguishably different chemical composition (see Fig.~\ref{fig:XO2system_dXH_ZI}), with a mean elemental abundance 
difference between the N and the S components of $+0.067 \pm 0.032$ dex, and statistically significant trends with the condensation temperature 
(Fig.~\ref{fig:XO2system_dXH_Tc}). It is normally assumed that individual components in binary systems share the same origin and 
initial bulk abundances as they form within the same cloud. Therefore, under the reasonable assumption that the XO-2 components 
were born together from the same molecular cloud, their dissimilar elemental abundances seem to favor other explanations compared to 
Galactic chemical evolution and/or stellar birthplace effects as possible reasons of the abundance patterns we observe
(see, e.g., \citealt{adibekyanetal2014}).

Differences in elemental abundances between the XO-2 system components could then reflect the likely different formation and evolution histories of the 
two planetary systems orbiting around them, which might have led to metal deficiencies in XO-2S or excesses in XO-2N. We recall in fact the different 
architectures of the two planetary systems: XO-2N hosts a transiting hot Jupiter, with preliminary evidence for another object in an outer orbit 
(see Paper II, and references therein); XO-2S hosts two giant planetary companions on moderately close-in orbits (and possibly a third long-period object; see Paper I). 

We estimate the amounts of heavy elements being accreted or depleted in one of the components following the prescriptions
of \cite{murrayetal2001}, as done in \cite{desideraetal2004} and \cite{desideraetal2006}.
The mass of the convective envelope results to be about $0.06~M_{\odot}$. Considering the
stellar [Fe/H] and $\Delta$[Fe/H] between the components from Table \ref{tab:elemental_abundances}, it
results that the difference of iron within the convective zone is of $\sim 10~M_\oplus$.
Assuming to first order a similar composition of meteoritic material, this corresponds to $50~M_\oplus$
of heavy elements. These amounts are larger than those considered in similar studies on planetary accretion 
(e.g., \citealt{pinsonneaultetal2001}, \citealt{macketal2014}, \citealt{tuccimaiaetal2014}), 
due to the size of the convective envelope and the super-solar metallicity of the XO-2 components.
We consider in the following sub-sections the two alternatives of heavy elements depletion in XO-2S and heavy elements excess in XO-2N.

\subsection{Heavy elements depletion in XO-2S?}

The existence of the abundance difference between the components of the XO-2 system can be interpreted
in the context of the scenario devised by \cite{melendezetal2009} as depletion of heavy elements
in the stellar atmosphere of XO-2S, due to their being locked in planetary companions.
More precisely, our result appears qualitatively similar to that recently found by \cite{tuccimaiaetal2014} 
in the binary system 16 Cyg\,A/B, in which the secondary component, which hosts a giant planet in eccentric orbit,
is metal deficient with respect to the companion, which is not known to have planetary companions.
Moreover, as found by the same authors, here we reveal a hint of the break in $T_{\rm C}$ between volatiles and refractories at 
$\sim 800$\,K (Fig.~\ref{fig:XO2system_dXH_Tc}). As pointed out by \cite{tuccimaiaetal2014}, this means that the 
rocky core of the giant planet most probably was not formed in the inner disk, but at larger distances where they are more likely to form.

The large amounts of heavy elements to be locked in planetary companions is challenging to estimate. 
However, it is a possibility that can not be dismissed in a super metal-rich environment such as that of the XO-2 system.
Indeed, there are cases of compact transiting giant planets whose high density appears to be due
to a very large content of heavy elements, such as the hot-Saturn HD149026b (\citealt{sato2005}, \citealt{fortney06}).

Theoretical models also predict the existence of such high-density Saturn-mass planets formed by up to 80\% of
heavy elements (\citealt{mordasini12}), however they are most likely to form in the outer region of a super metal-rich 
protoplanetary disk (\citealt{dodsonrobinson2009}), so that substantial migration would have been needed to carry 
such a planet at the observed separation of XO-2Sb.

The inner planet XO-2Sb has a mass similar to HD149026b and stellar metallicities are also comparable.
Unfortunately, search for transits of XO-2Sb have been inconclusive up to now (APACHE Team, priv. comm.),
preventing the determination of the planet density.
The presence in XO-2S of two planets (with XO-2Sc more massive than XO-2Nb) and the possibility of an additional 
object responsible for the long term RV trend could further contribute to the segregation of heavy elements by the planets, 
favoring the locking scenario.
The moderate eccentricities of the XO-2Sb and  XO-2Sc ($e \sim 0.2$) could imply a smooth migration history 
of the known giant planets around XO-2S,
possibly different from a dynamical instability event that might have caused the inward migration of XO-2Nb to its 
current tight orbit. Such a smooth evolution could also have allowed formation
and survival of additional low-mass planets that remain undetected in the available RV data.
This possibility was also mentioned by \cite{teskeetal2015}.

\subsection{Heavy elements excess in XO-2N?}

The enrichment of refractory elements in XO-2N when compared with XO-2S could be due to accretion of material onto the 
star and subsequent ingestion of rocky material coming from already formed planets or from the inner part of the disk and pushed into 
XO-2N by its hot Jupiter as it migrated inward to the present position, most likely as a result of planet-planet scattering (\citealt{gonzalez1997, 
idalin2008, raymondetal2011}). Indeed in Paper II, it was discussed that migration to the current close orbit of XO-2Nb
can not be due to Kozai interaction for the present orbit configuration, supporting a phase of dynamical instability
in the planetary system. The almost null eccentricity and the low spin-orbit angle of XO-2Nb (see Paper II, and references therein)
are not relevant for inferences on the original system configuration and migration history, as they are expected to be due
to tidal effects considering the planet characteristics, the size of the convective envelope of the star, and the system age.

The dynamical instability could have been triggered by modifications of the binary orbit caused, e.g., by stellar encounter 
in a dense stellar environment or a former hierarchical triple system that became unstable after the planets completed their 
growth around the primary star (\citealt{marzaribarbieri2007}). 
Indeed, \cite{kaibetal2013} studied very wide binaries, indicating that they may reshape the planetary 
systems they host, changing the orbital eccentricity or causing the ejection of planets or favoring the ingestion of material 
by the host star. In particular, they demonstrated that in binary systems with stellar masses and mean separation similar 
to those of XO-2N/S, 90\% of instabilities occur after $\sim 100$\,Myr, i.e. after planet formation is complete. 
This might cause ingestion of planetesimals and/or planet ejections of one or more objects. 
This scenario has the advantage that the dynamical instability and the successive system reconfiguration could have occurred
well after the shrinking of the outer convective zone of the star after the pre-main sequence phase, then avoiding one
of the difficulties in the planetary accretion scenario.

A challenge can be represented by the different situation of the HAT-P-1 system, whose secondary also hosts a hot-Jupiter
with mass of $0.53 M_{\rm J}$ and orbital period 4.46529 days (\citealt{bakosetal2007}), so not markedly different from XO-2Nb.
The masses of the components of the system are slightly larger than those of the XO-2 system ($1.16 M_\odot$ and $1.12 M_\odot$)
and the binary separation is also large (1550 AU) as for the XO-2 system.
However, no significant abundance difference between the components have been reported by \cite{liuetal2014}.
It was suggested that while the smaller size of the convective envelope could make easier the imprinting of any planet signature 
in the system, the larger stellar mass favored a faster system evolution, with accretion before the shrinking of the 
convective zone, as more massive stars seem to have shorter disk lifetime (\citealt{williamscieza2011}). 
More likely, the stochastic nature of the dynamical instability events should imply different outcomes for what concerns 
the ingestion of already formed planets and planetary debris.

The amount of heavy elements to be ingested in the XO-2N enrichment scenario obviously corresponds to that assumed to be locked into planetary
bodies in the XO-2S depletion scenario. As shown above, low-mass giant planets could have heavy elements content of several tens of 
Earth masses, making the ingestion of an individual planet with peculiar characteristics a possible cause for the whole abundance difference.

\section{Conclusions}
\label{sec:conclusions}

We performed an accurate differential analysis of abundances of iron-peak, $\alpha$-, $s$-process, $r$-process, and other elements 
for the dual-planet-hosting binary XO-2, in which the stars show very similar stellar parameters ($T_{\rm eff} \sim 5300$~K, 
$\log{g}\sim 4.43$, $M{_\star} \sim 0.97 M_\odot$, $R_\star \sim 1.0 R_\odot$, and [Fe/H]$\sim 0.35$ dex). To our 
knowledge, this is the first binary system in which both components host planets and where elemental abundance differences 
were revealed (see also \citealt{teskeetal2015,ramirezetal2015}).

We find significative abundance differences in almost all 25 elements, in the sense that the slightly cooler component XO-2N is also the star 
with higher content of elemental abundances, with stronger enhancement in the refractories. The abundance differences range from $\sim 0$ dex 
for the volatiles up to $\sim 0.11$ dex for the refractory elements, with a significance at the $2\sigma$ level for almost all elements. 

The differential abundances of XO-2N relative XO-2S show positive trends with $T_{\rm C}$, with a significance of the slope 
at the $5\sigma$ level considering all elements. This could mean that both components 
experienced the accretion of rocky planetary interior to the subsequent giant planets. We discuss about the possibilities that this result 
could be interpreted as the signature of abundances deficiencies in XO-2S or excesses in XO-2N, i.e. due to the ingestion of material by XO-2N 
or depletion by locking of heavy elements by the XO-2S planets, respectively. We estimate a mass of several tens of Earth as amount 
in heavy elements being accreted/depleted. For the time being, we do not know which of these two scenarios dominates or which one is the 
most appropriate and conclude that both are plausible. Future detailed numerical simulations and theoretical interpretations will be needed to 
try to disentangle the skein and clarify whether one of the two hyphoteses is most probable or if both are at work in such system. This is out
of the aim of this work, which was finalized to look at possible differences in elemental abundances in a planet hosting binary.

We confirm the potentiality of the binary hosting stars as useful tools to derive in a very accurate way 
elemental abundance differences. Further detailed studies on chemical composition of double systems hosting 
planets can provide important constraints on stellar/planetary formation and subsequent evolution. Studies 
on how planet formation may influence the chemical composition of hosting stars can be important to select 
future exoplanet searches.

\begin{acknowledgements}
This Italian GAPS project acknowledges support from INAF through the ``Progetti Premiali'' funding scheme 
of the Italian Ministry of Education, University, and Research. We thank the TNG staff for help with the 
observations. K.B. also thanks V. D'Orazi and C. Sneden for fruitful discussions about heavy elements.
\end{acknowledgements}

\bibliographystyle{../../aa-package_Apr2013/aa}

\end{document}